\documentclass[11pt]{article}
\newcommand{\la}[1]{\label{#1}}
\newcommand{\be}{\begin{equation}}
\newcommand{\ee}{\end{equation}}
\newcommand{\ba}{\begin{eqnarray}}
\newcommand{\ea}{\end{eqnarray}}
\newcommand{\rmi}[1]{{\mbox{\scriptsize #1}}}

\newcommand{\eq}{Eq.~}
\newcommand{\se}{Sec.~}
\newcommand{\eqs}{Eqs.~}
\newcommand{\nr}[1]{(\ref{#1})}
\newcommand{\tr}{{\rm Tr\,}}
\newcommand{\im}{\mathop{\rm Im}}
\newcommand{\nn}{\nonumber \\}
\newcommand{\fr}[2]{{\frac{#1}{#2}\,}}

\renewcommand{\vec}[1]{{\bf #1}}

\newcommand{\Nf}{n_{\rm G}}
\newcommand{\Ns}{n_{\rm S}}

\newcommand{\rmO}{{\mathcal{O}}}

\def\lsi{\raise0.3ex\hbox{$<$\kern-0.75em\raise-1.1ex\hbox{$\sim$}}}
\def\gsi{\raise0.3ex\hbox{$>$\kern-0.75em\raise-1.1ex\hbox{$\sim$}}}
\newcommand{\lsim}{\mathop{\lsi}}

\newcommand{\Tint}[1]{{\hbox{$\sum$}\!\!\!\!\!\!\int}_{\!\!\!\!#1}}
\newcommand{\rmii}[1]{{\mbox{\tiny\rm{#1}}}}

\makeatletter \@addtoreset{equation}{section} \makeatother
\renewcommand{\theequation}{\arabic{section}.\arabic{equation}}
\makeatletter
\renewcommand\section{\@startsection {section}{1}{\z@}%
                                   {-5.5ex \@plus -1ex \@minus -.2ex}
                                   {2.3ex \@plus.2ex}%
                                   {\normalfont\large\bfseries}}
\renewcommand\subsection{\@startsection{subsection}{2}{\z@}%
                                     {-3.25ex\@plus -1ex \@minus -.2ex}%
                                     {1.5ex \@plus .2ex}%
                                     {\normalfont\normalsize\bfseries}}
\renewcommand\thesection {\@arabic\c@section}
\renewcommand\thesubsection   {\thesection.\@arabic\c@subsection}
\renewcommand{\@seccntformat}[1]{%
\csname the#1\endcsname.\hspace{1.0em}}
\makeatother

\begin{document}

\begin{titlepage}
\begin{flushright}
BI-TP 2005/27\\
hep-ph/0508195\\
\end{flushright}
\begin{centering}
\vfill

{\Large{\bf Real-time Chern-Simons term for hypermagnetic fields}}

\vspace{0.8cm}

M.~Laine 

\vspace{0.8cm}

{\em
Faculty of Physics, University of Bielefeld, 
D-33501 Bielefeld, Germany\\}

\vspace*{0.8cm}
 
\mbox{\bf Abstract}

\end{centering}

\vspace*{0.3cm}
 
\noindent
If non-vanishing chemical potentials are assigned to chiral
fermions, then a Chern-Simons term is induced for the corresponding 
gauge fields. In thermal equilibrium anomalous processes adjust 
the chemical potentials such that the coefficient of the Chern-Simons 
term vanishes, but it has been argued that there are non-equilibrium
epochs in cosmology where this is not the case and that,
consequently, certain fermionic number densities and large-scale 
(hypermagnetic) field strengths get coupled to each other. We generalise 
the Chern-Simons term to a real-time situation relevant for dynamical 
considerations, by deriving the anomalous Hard Thermal Loop effective 
action for the hypermagnetic fields, write down the corresponding 
equations of motion, and discuss some exponentially growing 
solutions thereof.
\vfill
\noindent
 

\vspace*{1cm}
 
\noindent
August 2005

\vfill

\end{titlepage}

%
\section{Introduction}

It was realised long ago that if chiral fermions are assigned
a non-zero chemical potential, then a Chern-Simons term appears
to be induced for the corresponding gauge fields~\cite{wr}.
This statement is not without ambiguities, however~\cite{ns}.
In fact a non-Abelian Chern-Simons term transforms non-trivially
under large gauge transformations, and therefore the induced
theory is well-defined only for certain imaginary 
values of the chemical potential~\cite{djt}. Formally, 
the reason for these problems is that the chiral charge is 
not conserved because of the axial anomaly~\cite{gth}, so that 
strictly speaking no chemical potential should be assigned to it. 

In the standard electroweak theory, chiral fermions couple
not only to non-Abelian gauge fields, but also to the Abelian  
hypercharge fields. Hence, at the high temperatures where
the electroweak symmetry is restored, a Chern-Simons term appears
to be induced for them as well. An Abelian Chern-Simons action does
not have the same topological properties as 
the non-Abelian one, but on the other hand it is gauge-invariant, 
and could thus conceivably have a more direct physical significance 
than its non-Abelian counterpart. 
 
Indeed, there have been a number of suggestions for possible 
roles that the Abelian hypermagnetic Chern-Simons term might
play in cosmology. One of them is related to the observation
that right-handed electrons, which do not take part
in weak interactions and
also have a very small Yukawa coupling, are practically decoupled from 
the thermal ensemble above temperatures of about 10 TeV~\cite{cdeo}.
If they come with a non-vanishing net density, which can be 
described by a chemical potential, then a hypermagnetic
Chern-Simons term gets induced. It has been argued that this
leads to an instability and to the subsequent generation of 
large-scale hypermagnetic fields~\cite{js}. While it is 
believed that any length scales related to physics within 
the horizon of this epoch are too small to act as seeds for 
the currently observed galactic magnetic fields~\cite{decay}, 
such fields could have other physical consequences, for 
instance affecting the properties of the electroweak phase 
transition~\cite{ewpt}, the sphaleron energy~\cite{gr},
and electroweak baryogenesis~\cite{pa} 
(in suitable extensions of the Standard Model).

Another possible role acts in the opposite direction. Suppose
that there exist primordial hypermagnetic fields as a result  
for instance of some inflationary dynamics. Then anomalous 
processes could convert some of these fields to lepton and
eventually to baryon number, resulting possibly  
in the existence of matter and antimatter domains~\cite{gs}, 
which could lead to nucleosynthesis taking place in 
a corresponding environment~\cite{kss}. 

Whichever of these physics effects is realised, 
a central ingredient is always the presence of a  
hypermagnetic Chern-Simons term induced by a fermionic 
chemical potential. As argued in Refs.~\cite{js,gs}, the Chern-Simons 
term leads to an additional ``anomalous'' term in the magnetohydrodynamic
equations that govern the evolution of the hypermagnetic fields.   
The goal of this paper is to attempt a field theoretic derivation
of the equations of motion for the hypermagnetic fields in the 
presence of fermionic chemical potentials. The appropriate
framework is that of the Hard Thermal Loop (HTL) effective 
theories~\cite{htl,htlaction}. 
By integrating out the ``hard'' fermions, 
with energies of the order of the temperature $T$, we derive the 
effective action for the ``soft'' gauge fields, with wave vectors $\vec{p}$
and Minkowskian frequencies $\omega$ much smaller than the 
temperature, $|\vec{p}|, |\omega| \ll 2\pi T$. The standard static
Chern-Simons term used for instance in the considerations
of Refs.~\cite{js,gs} is recovered if we make the further
approximation $|\omega| \ll |\vec{p}|$ (which indeed 
appears to be well justified in practice).

The plan of this note is the following. 
In \se\ref{se:static} we recapitulate the static
gauge field effective action at high temperatures, and
in \se\ref{se:htl} generalise it to the non-static situation. 
We solve the resulting equations of motion for a simple 
case in~\se\ref{se:soln}, and conclude in~\se\ref{se:concl}.

%
\section{Anomalous effective action in the static limit}
\la{se:static}

Let us consider the standard electroweak theory at temperatures $T$
above a few hundred GeV. Let $A^a_\mu$, $B_\mu$ be the SU(2)$_\rmi{L}$ 
and U(1)$_\rmi{Y}$ gauge fields, $G^a_{\mu\nu}, F_{\mu\nu}$ the 
corresponding field strength tensors, and $g,g'$ the gauge couplings. 
The covariant derivative reads
$
 D_\mu = \partial_\mu - i g T^a A^a_\mu + i g' \frac{Y}{2} B_\mu
 \;, 
$
where $T^a$ are Hermitean generators normalised as 
$\tr[T^aT^b] = \delta^{ab}/2$, and $Y$ is the hypercharge
quantum number. With these conventions, the gauge field part
of the dimensionally reduced Euclidean Lagrangian~\cite{dr}  
takes the form
\ba
 \mathcal{L}_E & = & 
   f_E
 + \fr14 G^a_{\mu\nu}G^a_{\mu\nu} + \fr14 F_{\mu\nu}F_{\mu\nu}
 + \fr12 m_E^2 A_0^a A_0^a + \fr12 m_E'^2 B_0 B_0 + i j_E' B_0 + 
 \nn & + & 
 c_E\, n_\rmi{CS} + c_E'\, n_\rmi{CS}' 
 + ...
 \;, \la{static}
\ea
where the ``anomalous'' Chern-Simons densities read
\ba
 n_\rmi{CS}  & \equiv & \frac{g^2}{32\pi^2} 
  \epsilon_{ijk}
  \Bigl(
   A^a_i G^a_{jk} - \frac{g}{3} f^{abc} A^a_i A^b_j A^c_k 
  \Bigr)	
 \;, 
 \la{NCS} \\ 
 n_\rmi{CS}' & \equiv & 
 \frac{g'^2}{32\pi^2}
  \epsilon_{ijk}
  \, B_i F_{jk}   
 \;, \la{NCSp}
\ea
and infinitely many
higher dimensional operators have been suppressed.
The term $f_E$ is a (field-independent) ``unit operator''. The relative 
parametric
accuracy that can be reached with dimensionally reduced effective
theories of this kind has been analysed in Ref.~\cite{parity}.

There are a number of matching coefficients appearing in \eq\nr{static}.
Denoting by $\Nf$ the number of generations, 
by $\Ns$ the number of fundamental scalar doublets, 
by $\mu_Q$ the common chemical potential of the quarks, and 
by $\mu_{L_i}$ ($\mu_{R_i}$) the chemical potential of the $i^\rmi{th}$
left-handed (right-handed) lepton generation, 
the leading-order expressions read
\ba
 f_E & = & 
 -\biggl(24 + \frac{105}{4}\Nf + 4 \Ns \biggr) \frac{\pi^2T^4}{90}
 -\biggl( \Nf \mu_Q^2 + 
 \sum_{i=1}^{\Nf} \frac{2 \mu_{L_i}^2 + \mu_{R_i}^2}{12}
  \biggr)T^2 
 - \nn & & 
 - \Nf \frac{\mu_Q^4}{2\pi^2} 
 - \sum_{i=1}^{\Nf} \frac{2 \mu_{L_i}^4 + \mu_{R_i}^4}{24\pi^2}
 \;, \la{fE} \\ 
 m_E^2 & = & g^2 \biggl[ 
 \biggl(
  \fr23 + \frac{\Nf}{3} + \frac{\Ns}{6} 
 \biggr) T^2 + 3 \Nf \frac{\mu_Q^2}{4\pi^2}
 + \sum_{i=1}^{\Nf} \frac{\mu_{L_i}^2}{4\pi^2}
 \biggr]
 \;, \la{mmE} \\
 m_E'^2 & = & g'^2 \biggl[ 
  \biggl( \frac{5\Nf}{9} + \frac{\Ns}{6}
  \biggr) T^2
 + 11 \Nf \frac{\mu_Q^2}{12\pi^2}
 + \sum_{i=1}^{\Nf} \frac{\mu_{L_i}^2 + 2 \mu_{R_i}^2}{4\pi^2}
 \biggr]
 \;, \la{mmEp} \\
 j_E' & = & g' 
 \biggl[
   \frac{\Nf \mu_Q}{3}
   \biggl( T^2 + \frac{\mu_Q^2}{\pi^2} 
   \biggr) - 
   \sum_{i=1}^{\Nf}
   \biggl( 
    \frac{\mu_{L_i}+\mu_{R_i}}{6} T^2
    + \frac{\mu_{L_i}^3 + \mu_{R_i}^3}{6\pi^2}
   \biggr)
 \biggr]
 \;, \la{jE} \\ 
 c_E & = & 3 \Nf \mu_Q^{\mbox{ }} 
 + \sum_{i=1}^{\Nf}\mu_{L_i}^{\mbox{ }}
 \;, \la{cE} \\ 
 c_E' & = & -c_E + 2 \sum_{i=1}^{\Nf} 
 \Bigl( \mu_{L_i}^{\mbox{ }} - \mu_{R_i}^{\mbox{ }} \Bigr)
 \;. \la{cEp}
\ea
Some higher-order corrections can be found in Ref.~\cite{ag}. 

Now, because of the axial anomaly, 
the rate of baryon plus lepton number violation, 
${\rm d} \ln |B+L| / {\rm d}t \approx -(13\Nf/4) (25.4\pm 2.0) \alpha_w^5 T$%
~\cite{asy}--\cite{gdm},\footnote{
 The number 25.4 is in fact the value 
 of a function containing terms like $\ln(1/\alpha_w)$, 
 at the physical $\alpha_w$.
 } 
is significantly larger than the expansion rate of the Universe, 
$\sim T^2/m_\rmi{Pl}$, for $10^2$~GeV $\lsim T \lsim$ $10^{12}$~GeV, 
so that the anomalous processes are 
perfectly in thermal equilibrium~\cite{krs}.
Therefore, the corresponding chemical potential 
should be set to zero: 
\be 
 \mu_{B+L}^{\mbox{ }} \equiv 3 \Nf \mu_Q^{\mbox{ }} 
 + \sum_{i=1}^{\Nf}\mu_{L_i}^{\mbox{ }} = 0 
 \;. 
\ee
In other words, the coefficient $c_E$ in \eq\nr{cE} vanishes.

On the other hand, the coefficient $c_E'$ in \eq\nr{cEp}
does not vanish, provided that $\mu_{R_i} \neq \mu_{L_i}$. 
Such a situation can arise if chirality flipping processes, 
mediated by the Yukawa couplings, are out of equilibrium~\cite{cdeo},
and we will assume this to be the case in the following. Formally, 
this can be reached by setting the electron Yukawa coupling to zero. 

The naive conversion of \eq\nr{static} to Minkowski
spacetime goes simply through the analytic continuation
\be
 \partial^E_0 = - i \partial^M_0 \;, \quad
 A^{aE}_0 = -i A^{aM}_0 \;, \quad
 B^E_0 = -i B^M_0 \;, \quad
 \mathcal{L}_E = - \mathcal{L}_M 
 \;,  \la{analytic}
\ee
and the Minkowskian action is then given by 
$S_M = \int \! {\rm d}t\, {\rm d}^3 x \, \mathcal{L}_M$.
The resulting theory is gauge invariant only in 
static gauge transformations, however, and thus cannot be
the full truth. In the next Section, we recall
how a more precise theory can be obtained.

%
\section{Anomalous Hard Thermal Loop effective action}
\la{se:htl}

Suppose that we stay for a further moment in the static limit, 
and consider what kind of higher order operators could appear
in \eq\nr{static}. From the point of view of the original
four-dimensional theory, some of these operators arise 
from a gradient expansion in spatial derivatives. Given that
the scale that has been integrated out to obtain \eq\nr{static}
is the ``hard'' scale $\sim 2 \pi T$, these 
operators are necessarily suppressed by (at least) 
$\rmO( |\mathbf{\nabla}|^2/(2\pi T)^2 )$ with respect to the ones 
that have been kept in \eq\nr{static}.

Now, one might expect that the same is true for 
temporal derivatives: maybe their effects are also suppressed
by $\partial_0^2/(2\pi T)^2$? This is not the case! 
As is well-known from Hard Thermal Loop
considerations~\cite{htlold,htl,htlaction}, 
time-dependence is suppressed with respect 
to the static limit only by $\rmO(|\partial_0| / |\nabla|)$, and is
in general of order unity. 

Let us proceed with the explicit computation. 
We employ the Matsubara 
formalism, followed by analytic continuation. 
The chemical potential corresponds in momentum space 
to shifting fermionic Matsubara frequencies $\omega_n$ 
as $\omega_n \to \omega_n + i\mu$. We use the
(sometimes implicit) notation that summing over Lorentz indices 
which are both down implies the use of Euclidean metric.
Capital momenta ($P,Q,R$) are assumed Euclidean.

With these conventions, 
for any given left-handed fermion with hypercharge $Y$, the 
anomalous part of the hypermagnetic
Euclidean action in momentum space reads
\ba
 \delta S_{E} & = &  \fr14 g'^2 Y^2 
 \Tint{\,Q,R} \delta_{Q+R}\, 
 B_\mu(R) B_\nu(Q) \Gamma_{\mu\nu}(Q) 
 \;, \\ 
 \Gamma_{\mu\nu}(Q)  & \equiv & 
 \Tint{P=(\omega_n + i \mu,\vec{p})}
 \frac{\Delta_{\mu\nu}(p_0,\vec{p})}
 {(P+Q)^2 P^2}
 \;, \la{SE3} \\
 \Delta_{\mu\nu}(p_0,\vec{p}) & \equiv &  
 \fr14 \tr [\,\gamma_\mu (\slash\!\!\!\! P + \slash\!\!\!\! Q) 
 \gamma_\nu \,\slash\!\!\!\! P \gamma_5 ]
 \;, \la{gamma} 
\ea
where $\raise-0.4mm\hbox{$\Sigma$}\!\!\!\!\!\int\,$ 
denotes the usual Matsubara sum-integral
(bosonic for $Q,R$ and fermionic for $P$). 
For right-handed fermions the overall sign is opposite.
The Euclidean $\gamma$-matrices here are Hermitean, with 
$\{\gamma_\mu,\gamma_\nu\} = 2 \delta_{\mu\nu}$, and we have 
defined $\gamma_5 \equiv \gamma_0 \gamma_1 \gamma_2 \gamma_3$. 
It is perhaps appropriate to mention that
it is possible to have an ``anomalous'' term with
two gauge field legs only, since the chemical potential 
effectively acts as a third leg.

To evaluate the sum over $\omega_n$
in \eq\nr{SE3}, we note 
that there are only single poles in $p_0$ and 
that we can thus use the contour formula
\be
 T \sum_{n \rmi{ odd}}f(n \pi T + i \mu) = 
 \int_{-\infty}^{\infty}
 \! \frac{{\rm d} p_0}{2\pi} f(p_0) 
 + \sum_\rmi{$\im z < 0$} 
 \frac{i \mathop{\mbox{Res}}[ f(z) ]}{e^{i \beta z + \beta \mu} + 1}
 - \sum_\rmi{$\im z > 0$} 
 \frac{i \mathop{\mbox{Res}}[ f(z) ]}{e^{-i \beta z - \beta \mu} + 1}
 \;, \la{res}
\ee
where the sums are over the poles $z$ of $f(z)$. The first term 
is independent of $T$ and $\mu$; 
we ignore this zero-temperature vacuum part here. The latter
two terms are ultraviolet and infrared finite, and require
no regularization. Thus we can work in exactly four dimensions
(as already implicitly assumed above), 
whereby $\tr [\gamma_5 \gamma_\mu\gamma_\nu\gamma_\alpha\gamma_\beta] = 
4 \epsilon_{\mu\nu\alpha\beta}$, with $\epsilon_{0123} = +1$.

Picking up the four poles; shifting integration variables in 
three of them\footnote{%
 Shifting integration variables is safe provided
 that each individual term is finite, as is the case here.
 We have however
 checked the outcome also by not carrying out any shifts but 
 just expanding in $Q/\omega_\vec{p}$.
 } 
as $\vec{p}\to -\vec{p}$, 
$\vec{p}\to \vec{p} - \vec{q}$, 
$\vec{p} \to -\vec{p} - \vec{q}$; and employing the facts that 
$\Delta_{\mu\nu}(-p_0,-\vec{p}) = -\Delta_{\mu\nu}(p_0,\vec{p})$, 
$\Delta_{\mu\nu}(p_0-q_0,\vec{p}-\vec{q}) = \Delta_{\mu\nu}(p_0,\vec{p})$, 
$\Delta_{\mu\nu}(-p_0-q_0,-\vec{p}-\vec{q}) = -\Delta_{\mu\nu}(p_0,\vec{p})$; 
the thermal part of~\eq\nr{SE3} becomes    
\ba
 \Gamma_{\mu\nu}(Q) \!\! & = & \!\! 
 \int_\vec{p} \frac{N_-(\omega_\vec{p})}{4 \omega_\vec{p}^2}
  \Delta_{\mu\nu}(-i \omega_\vec{p},\vec{p})
 \biggl( 
   \frac{1}{v^E\cdot Q + \frac{Q^2}{2\omega_\vec{p}}} - 
   \frac{1}{v^E\cdot Q - \frac{Q^2}{2\omega_\vec{p}}}
 \biggr)
 \;, \la{s2}
\ea
where we have denoted $\int_\vec{p} \equiv \int \! {{\rm d}^3p}/{(2\pi)^3}$,
$N_-(\omega_\vec{p}) \equiv 
n_\rmi{F} (\omega_\vec{p}-\mu)- n_\rmi{F} (\omega_\vec{p}+\mu)$, 
$n_\rmi{F} (\omega_\vec{p}) \equiv {1}/({e^{\beta \omega_\vec{p}} + 1})$,
$v^{E}_\mu \equiv (-i,p_i/\omega_\vec{p})$, 
$\omega_\vec{p} \equiv |\vec{p}|$, 
and $v^E\cdot Q \equiv v^E_\mu Q_\mu$. 
Note that no approximations have been made so far
for the thermal part. 

The next step is to carry out a small coupling expansion.
In other words, we look for the leading term in the expansion in 
small $|Q|/\omega_\vec{p}$, where parametrically (after the analytic 
continuation to follow presently) $Q$ is a soft scale, 
$|Q| \lsim {\rm max}(gT,g\mu)$, 
while the integration variable gets its contributions from 
the hard scales, 
$\omega_\vec{p}\sim {\rm max}(T,\mu)$.
The leading term in the expansion of the denominators
in \eq\nr{s2} obviously cancels, but the next-to-leading 
term is non-vanishing, and multiplied by 
$\Delta_{\mu\nu}(-i\omega_\vec{p},\vec{p}) = 
\omega_\vec{p} \epsilon_{\mu\alpha\nu\beta} Q_\alpha v^E_\beta$.
Thus, we obtain 
\ba
 \Gamma_{\mu\nu}(Q) & \approx & 
 \epsilon_{\mu\nu\alpha\beta}\,
 Q^2 
 \int_\vec{p} \frac{N_-(\omega_\vec{p})}{4\omega_\vec{p}^2}
 \frac{Q_\alpha v^E_\beta}{(v^E\cdot Q)^2}
 \;. \la{appG}
\ea

At this point we carry out analytic continuation to Minkowski 
spacetime. Any Euclidean Lorentz-vector can be written as  
$f^E_\alpha = \Lambda_{\alpha\beta} f^{M\beta}$, with 
$\Lambda_{\alpha\beta} = \mathop{\mbox{diag}}(-i,-1,-1,-1)$.
Furthermore, as mentioned
in \eq\nr{analytic}, there is an overall minus-sign between 
$\mathcal{L}_E$ and $\mathcal{L}_M$. 
We also introduce an angular integration 
$\int_v \equiv  \int \! {{\rm d} \Omega_v}/ {4\pi}$, 
where $v^\mu = (1, v^i)$, 
$v_\mu v^\mu = 0$, and
the integral is over the directions $\Omega_v$ of $v^i$, 
and note that the radial integration in~\eq\nr{appG}
can be carried out exactly, with the result 
\be
 \int_\vec{p} \frac{N_-(\omega_\vec{p})}{\omega_\vec{p}^2}
 = 
 - \int_\vec{p} 
 \frac{N'_-(\omega_\vec{p})}{\omega_\vec{p}} 
 = \frac{\mu}{2\pi^2}
 \;. \la{int3}
\ee

Summing over all fermions, 
adding the known non-anomalous
bosonic terms~\cite{htlaction},
and going furthermore to $x$-space, 
we arrive at our final action:
\ba
 S_M \!\! & = & \!\!
 \int_x \biggl[ -\fr14 F_{\mu\nu}F^{\mu\nu} \biggr] + S_\rmi{HTL}
 \;, \la{Sfree} \\
 S_\rmi{HTL} \!\! & = & \!\! \int_{x,v} 
 \biggl[
 - j_E' v^\mu B_\mu   
 - \frac{m_E'^2}{4}  
   F_{\alpha\mu} 
   \frac{v^\alpha v^\beta}{(v\cdot\partial)^2} 
   {F_\beta}^\mu
 + c_E'  \frac{g'^2}{32\pi^2}
   \widetilde F_{\alpha\mu}(x) \frac{v^\alpha}{(v\cdot\partial)^2}
   \partial^2 B^\mu(x)  
 \biggr]
 \;,  \hspace*{0.5cm} \la{SHTL}
\ea
where $\widetilde F_{\alpha\mu} \equiv 
\epsilon_{\alpha\mu\beta\nu} F^{\beta\nu}$
and $\partial^2 \equiv \partial\cdot\partial \equiv \partial^\mu\partial_\mu$.
The coefficients $m_E'^2, j_E', c_E'$
are the same as in \eqs\nr{mmEp}, \nr{jE}, and \nr{cEp}.%
  \footnote{%
  We note that a single chiral fermion contributes to the parameters as 
  $j_E' = g' \mu Y (T^2 + \mu^2/\pi^2)/12$,
  $m_E'^2 = g'^2 Y^2 (T^2/3 + \mu^2/\pi^2)/8$,
  $c_E' = - \mu Y^2 H /2$, 
  where $H = +1$ ($-1$) for right-handed (left-handed) fermions.}
The integrals $\int_v(...)$ are collected in Appendix~A. 

It is the last term in \eq\nr{SHTL} which is the new one. 
For more symmetry, one could replace 
$B^\mu \to v^\beta/(v\cdot \partial) {F_{\beta}}^\mu$ in it,
since the extra term $-[\partial^\mu / (v\cdot \partial)] v\cdot B$
thus introduced vanishes after partial integration and neglectance
of surface terms.

Note that \eq\nr{SHTL} is gauge invariant in time-dependent gauge 
transformations, unlike \eq\nr{static}. Nevertheless, 
by employing the identities in \eqs\nr{pi1}, \nr{pi2},
it is straightforward to rewrite $S_M$ in a form where it is obvious that
the hypercharge part of \eq\nr{static} is recovered in the static limit,
up to corrections of order $\rmO(|\partial_0|/|\nabla|)$. 

For completeness, we remark that although we have shown 
in~\eq\nr{SHTL} only the terms with the three 
largest coefficients, 
$j_E' \sim \rmO(\mu T^2), m_E'^2\sim \rmO(T^2)$, and $c_E'\sim \rmO(\mu)$, 
some additional operators
are known as well: there is a non-anomalous but 
charge-conjugation violating ``cubic'' 
operator with a coefficient $\rmO(\mu)$~\cite{htlmu}, and 
it appears that even ``quartic'' operators with a 
coefficient $\rmO(1)$ could be added, 
by reformulating the HTL action in terms of 
a certain matrix-valued kinetic theory~\cite{cm}.

Since the action in \eqs\nr{Sfree}, \nr{SHTL} is quadratic in the fields, 
its contents can equivalently be expressed through 
equations of motion.  It is useful to express
the equations of motion 
in a form following from the identities
in \eqs\nr{pi1}, \nr{pi2}. We obtain
\ba
 \partial^\mu F_{\mu\nu}(x) & = & 
 \int_v \biggl[ 
 j_E' v_\nu 
 - m_E'^2 \frac{v_\nu v^\alpha}{v\cdot\partial} F_{\alpha 0}(x)
 - c_E' \epsilon_{0\nu\alpha\beta} \frac{g'^2}{8\pi^2}
  \frac{v^\alpha}{v\cdot\partial} \partial^2 B^\beta(x)
 \biggr]
 \;. \la{eom}
\ea
The integrals have 
the properties $\int_v v^\alpha = {\delta^\alpha}_0$
and $(1 - {\delta^\alpha}_0)\int_v v^\alpha / v \cdot \partial 
\propto \partial^\alpha$ (cf.\ Appendix~A), which imply that 
the right-hand side is divergenceless (or transverse)
with respect to $\partial^\nu$. 

We should like to stress 
already at this point that there are circumstances
where higher order corrections to \eq\nr{eom} 
can become important. Let us recall, 
to start with, that standard HTL structures like the second
term in~\eq\nr{eom} can be reproduced~\cite{kin} by classical 
kinetic theory, or ``Vlasov equations'', of the type 
\ba
 && \Bigl[ 
  p \cdot \partial_x - \frac{g' Y}{2} p^\alpha F_{\alpha\beta}\partial_p^\beta
 \Bigr] f(x,p) \equiv  0
 \;, \la{vlasov} \\ 
 && \partial^\mu F_{\mu\nu}  \equiv  \frac{g' Y}{2} 
 \int_\vec{p}
 \int_{-\infty}^{\infty} \! \frac{{\rm d} p_0}{2\pi} 
 \, p_\nu \, f(x,p) 
 \;, \la{vlasov2}
\ea
by solving them perturbatively to second order in $g'$ 
with the ``initial condition'' that for $g'\to 0$ the solution reads
$f(x,p) \equiv f_{(0)}(x,p)\equiv 4\pi \delta(p^2) 
n_\rmi{F} (|p_0| - \mathop{\mbox{sign}}(p_0) \mu)$.
One may then expect higher order
interactions to generate a Boltzmann type
collision term on the right-hand side of \eq\nr{vlasov}. 
Indeed, effective theories which such a structure  
have recently been analysed in Ref.~\cite{eff}, and 
the collision term does turn out to be important in many contexts.
We will return to this issue presently.

Finally, the question arises whether the last term in~\eq\nr{eom}
can also be expressed through Vlasov equations. 
In principle this indeed is the case: for instance, 
we can ``by brute force'' generalise
$f(x,p)$ to a two-index Lorentz tensor, and the Vlasov equations to
\ba
 && \!\!
 \Bigl[
  p \cdot \partial_x - \frac{g' Y}{2} 
  p^\alpha F_{\alpha\beta} \partial^{\beta}_p 
 \Bigr] f^{\gamma\delta}(x,p)
  - \frac{g' Y}{2}
 {\epsilon^{\gamma\delta}}_{\alpha\rho} 
 \partial_x^2 B^\alpha \partial^p_\sigma f^{\rho\sigma}(x,p) \equiv 0 
 \;, \la{newvl1} \\
 && \!\!
  \partial^\mu F_{\mu\nu} \equiv
  \frac{g'Y}{2}\int_\vec{p} \! 
  \int_{-\infty}^\infty  \frac{{\rm d} p_0}{2\pi}
  \Bigl[
   \, \eta_{\gamma\delta} \, p_\nu + 
   H (\eta_{\gamma\nu}\, p_\delta - \eta_{\delta\nu}\, p_\gamma) 
  \Bigr] f^{\gamma\delta}(x,p)
 \;, \la{newvl2} \\
 && \!\!
  f^{\gamma\delta}_{(0)} (x,p) \equiv
  \eta^{\gamma\delta}\pi \delta(p^2)
  n_\rmi{F}(|p_0| - \mathop{\mbox{sign}}(p_0) \mu)
 \;,  
\ea
where $H = +1$ ($-1$) for right-handed (left-handed) fermions.
Solving this set perturbatively to second 
order in $g'$, it is easy to check
that \eq\nr{eom} is reproduced (with a single chiral
fermion contributing as specified in footnote~3). 
Nevertheless the new parts in these equations
are not particularly satisfactory: \eq\nr{newvl1} generates
a distribution which contains an ugly $\delta'(p^2)$ and  is 
not gauge invariant. While both problems disappear after the 
construction of the current in \eq\nr{newvl2}, it should be
possible to find a  more compelling formulation which also 
has a physically understandable interpretation.  

%
\section{On the growth rate of instabilities}
\la{se:soln}

In order to illustrate the effects that may originate
from the non-local HTL structures, we inspect the spatial part 
of \eq\nr{eom} (i.e.~$\nu = 1,2,3$). 
We employ a class of gauges where
$B_0$ is constant, so that 
$F_{k 0} = -\partial_0 B_k$, where $k$ is a spatial index.

We start by considering the static limit. Then 
\eq\nr{eom} becomes ($k = 1,2,3$)
\be
 [\nabla^2\delta_{ki}-\partial_k\partial_i ]B_i(x) = 
 c_E' \frac{g'^2}{16\pi^2} \epsilon_{kij} F_{ij}(x)
 \;. \la{ev}
\ee
Going into momentum space 
[$B_k(x) = \int_\vec{q} \tilde B_k(\vec{q})\exp(-i \vec{q}\cdot \vec{x})$], 
and choosing a frame where $\vec{q} = (0,0,q^3)$, we obtain
equations for the transverse components $\tilde B_1, \tilde B_2$.
They have a non-trivial solution provided that 
\be
 |\vec{q}| = \pm |q_\rmii{CS}| \;, \quad
 q_\rmii{CS} \equiv \frac{g'^2 c_E'}{8\pi^2} 
 \;. \la{qCS}
\ee
Going back to configuration space, the solution reads
\be
  B_1(x^3) = C \, \cos[q_\rmii{CS} (x^3 - x^3_0)] \;, \quad
  B_2(x^3) = C \, \sin[q_\rmii{CS} (x^3 - x^3_0)]
 \;,
\ee
where $C,x^3_0$ are constants.
This is nothing but the so-called (static) 
Chern-Simons wave~\cite{var}. 

Consider then the dynamical situation.
We transform \eq\nr{eom}
to Fourier space with respect to space coordinates
but keep the time coordinate in configuration space. 
Building on \eq\nr{ev}, we look for 
transverse modes which satisfy the eigenvalue equations ($k' = 1,2$)
\be
 \vec{q}^2 \tilde B_{k'}  + i q_\rmii{CS} \epsilon_{k'ij} q_i \tilde B_j
 = \lambda \tilde B_{k'}
 \;.  \la{ev2}
\ee
Two non-trivial solutions exist, with the eigenvalues 
\be
 \lambda = \vec{q}^2 \pm q_\rmii{CS}|\vec{q}|
 \;. 
\ee
Thus, for long wavelengths, $|\vec{q}| < |q_\rmii{CS}|$, there 
exists an unstable branch with $\lambda < 0$.

How fast do the modes with $\lambda < 0$ grow?
To find out, we assume that the time evolution is very 
slow, $|\partial_t^2| \ll |\lambda|$, and justify this assumption 
a posteriori. In this situation, we can approximate the complicated 
HTL structures through the leading terms in time derivatives, and 
the problem becomes tractable.  

Inspecting the integrals in Appendix A, it can be seen that both 
the 2nd and the 3rd term on the right-hand side of \eq\nr{eom} 
have a term linear in time derivatives. For a solution of 
\eq\nr{ev2}, however, the third term gives a contribution 
suppressed by 
$\rmO[(\vec{q}^2 - \lambda)/m_E'^2]\sim \rmO[q_\rmii{CS}^2/m_E'^2]$ 
with respect to the second term. This is very small for weak coupling, 
$\sim g'^2\mu^2/\mathop{\mbox{max}}(T^2,\mu^2)\pi^4$, 
and can safely be ignored. Thus the effect comes
from the 2nd term, and we obtain 
\be 
 - \lambda \tilde B_{k'} \approx 
 \frac{\pi}{4 |\vec{q}|} m_E'^2 \partial_t \tilde B_{k'}
 \;. \la{sigma}
\ee
To summarise, modes with $\lambda < 0$ grow exponentially,
$\tilde B_{k'} \sim\exp(\Gamma t)$, 
with the rate 
\be
 \Gamma \approx  \frac{4|\lambda \vec{q}|}{\pi m_E'^2} \sim 
 \frac{q_\rmii{CS}^2}{m_E'^2} |\vec{q}| 
 \;. \la{Gamma}
\ee
Given that $q_\rmii{CS}^2 / m_E'^2 \ll 1$, 
the assumption of slow growth ($\Gamma^2 \ll \vec{q}^2$) 
is indeed justified. 

It is important to realise at this point, 
though, that higher order corrections may give large 
contributions on the right-hand side of \eq\nr{sigma}, 
as already mentioned above.
In fact, it could happen that
summing an infinite set of higher order loop contributions
effectively shields the scale $|\vec{q}|$ in the denominator
of $\pi m_E'^2/4|\vec{q}|$ 
by a constant $\sim g'^4 \ln(1/g') T$, whereby the right-hand 
side goes over into $\sigma' \partial_t \tilde B_{k'}$, where 
$\sigma'$ is the hyperelectric conductivity, 
$\sigma' \sim T/g'^2 \ln(1/g')$~\cite{amy}.
Another way to think of the issue\footnote{%
 We stress that the discussion here is only
 qualitative in nature, and omits important points.
 } 
is that 
interactions may generate a ``thermal width'' $\Gamma_\rmi{th}$ 
of order $\Gamma_\rmi{th}\sim g'^4 \ln(1/g') T$ for the hard on-shell 
particles~\cite{ga} and that, if $q_0 \to q_0 + i \Gamma_\rmi{th}$ and
$|\vec{q}| \ll \Gamma_\rmi{th}$ in \eq\nr{L}, then 
the static limit would become $-i/\Gamma_\rmi{th}$ rather
than $-i\pi/2 |\vec{q}|$, amounting to the same shielding. 
Since $|\vec{q}|\sim q_\rmii{CS}\sim g'^2 \mu$
and $\Gamma_\rmi{th}\sim g'^4 \ln(1/g') T$,
we should formally assume $|\vec{q}| \gg \Gamma_\rmi{th}$
so that the shielding is irrelevant, 
but since typically $\mu \ll g'^2 T$ in cosmology,
the formal hierarchy may get reversed so that 
we indeed find ourselves in the situation 
$|\vec{q}| \ll \Gamma_\rmi{th}$. In fact, 
inserting numerical values relevant for the Standard Model, 
it appears that having
$|\sum_{i=1}^{\Nf}(\mu_{L_i}^{\mbox{ }} - \mu_{R_i}^{\mbox{ }})|/T \lsim 0.5$
already brings us to the reversed situation.

The estimates for the growth rate 
of hypermagnetic fields that were presented in Refs.~\cite{js,gs}
were based on the contribution of
the hyper\-electric conductivity $\sigma'$, 
rather than \eq\nr{sigma}, and should thus be correct 
under the phenomenologically relevant circumstances $\mu \ll T$. 
(The expansion of the Universe as well as the 
time-dependence of the chemical potentials, due to the backreaction
via the hypermagnetic part of the 
anomaly equation as well as the chirality flipping
processes induced by the electron Yukawa coupling, 
were also taken into account.)
We note that the corresponding 
growth rate is much {\em larger} than \eq\nr{Gamma}, 
since $\Gamma_\rmi{th} \gg |\vec{q}|$.

%
\section{Conclusions}
\la{se:concl}

We have addressed here the coupling between 
hypermagnetic fields and fermionic chemical potentials
in the standard electroweak theory at high temperatures.
This problem has phenomenological relevance in cosmology,
provided that a lepton asymmetry and/or primordial 
hypermagnetic fields exist at temperatures above the 
electroweak phase transition,
but is also related to some intriguing theoretical issues,
such as that the coupling discussed
seems to allow for a sharp distinction
between the high-temperature and low-temperature phases 
of the electroweak theory~\cite{allorder}.

Concretely, 
we have generalised the standard Abelian Chern-Simons term to an apparently 
Lorentz invariant form, which can be added to the Hard Thermal Loop 
action describing the real-time dynamics of the hypermagnetic 
fields (\eq\nr{SHTL}). 
We have also analysed 
the unstable exponentially growing solutions
that the resulting equations of motion have.
Our conclusion is that for such solutions,
the deviation of the anomalous term from its standard 
static form is in fact insignificant in practice
(cf.\ the discussion preceding \eq\nr{sigma}), 
so that the ignoring of this 
deviation in previous studies appears
well justified in retrospective.

The problem with the Hard Thermal Loop equations of motion
is that higher order corrections to their non-anomalous part
turn out to be very important for the small
values of chemical potentials that are assumed to appear
in cosmology, $\mu \ll T$.
(The wave vectors of the growing modes are proportional
to differences of chemical potentials and thus very
much smaller than the temperature in this situation.) 
In particular, the fact that a finite conductivity is 
expected to be generated through summing infinitely many 
high order loop corrections
modifies the growth rate of the unstable solutions significantly.
(The existence of unstable modes is not affected.)
Nevertheless, it seems to us that the conclusion mentioned above, 
namely that it is safe to use the static limit 
of the Chern-Simons term under 
phenomenologically relevant circumstances, 
continues to be valid. To be sure, it would of course
be interesting to develop a numerical framework where
both the Hard Thermal Loop effects discussed in this paper, and
higher order corrections such as conductivity, can be incorporated 
simultaneously, to allow for a more precise study of 
growing hypermagnetic fields.

Finally, we remark that the
growing hypermagnetic fields have a form which
generates a non-zero value for the hypermagnetic  topological charge 
density, $\sim g'^2F_{\mu\nu}\widetilde F_{\mu\nu}$. This means that 
fermion number densities and, consequently, chemical
potentials, evolve as dictated by the anomaly equation, 
in a way which stops the growth at some point. 
These processes have been analysed in Refs.~\cite{js,gs}.

%
\section*{Acknowledgements}

I wish to 
thank D.~B\"odeker and M.~Shaposhnikov for useful comments.

\newpage


\appendix
\renewcommand{\thesection}{Appendix~\Alph{section}}
\renewcommand{\thesubsection}{\Alph{section}.\arabic{subsection}}
\renewcommand{\theequation}{\Alph{section}.\arabic{equation}}


%
\section{Basic integrals}

Although well-known~\cite{htlold,htl,htlaction}, we collect here
the basic velocity integrals needed in this paper. Assuming implicitly that 
the frequency is replaced everywhere through $q_0\to q_0+i 0^+$, as is 
relevant for retarded Green's functions, the integrals read
($i,j=1,2,3$) 
\ba
 \int_v  & = & 1
 \;, \\
 \int_v v^i & = & 0
 \;, \\
 \int_v v^i v^j & = & \fr13 \delta^{ij}
 \;, \\  
 \int_v \frac{1}{v\cdot q} & = & L(q)
 \;, \\
 \int_v \frac{v^i}{v\cdot q} & = & \frac{q^i}{|\vec{q}|^2}
 \Bigl[ - 1 + q_0 L(q) \Bigr]
 \;, \\
 \int_v \frac{v^i v^j}{v\cdot q} & = & 
 \frac{L(q)}{2}
 \biggl( 
  \delta^{ij} - \frac{q^i q^j}{|\vec{q}|^2} 
 \biggr) + 
 \frac{q_0}{2 |\vec{q}|^2}
 \Bigl[ 1 - q_0 L(q) \Bigr]
 \biggl( 
  \delta^{ij} - 3 \frac{q^i q^j}{|\vec{q}|^2} 
 \biggr)  
 \;, \\  
 \int_v \frac{1}{(v\cdot q)^2} & = & \frac{1}{q^2}
 \;, \\
 \int_v \frac{v^i}{(v\cdot q)^2} & = & \frac{q^i}{|\vec{q}|^2}
 \Bigl[ \frac{q_0}{q^2} - L(q)  \Bigr]
 \;, \\
 \int_v \frac{v^i v^j}{(v\cdot q)^2} & = & 
 \frac{1}{2 q^2}
 \biggl( 
  \delta^{ij} - \frac{q^i q^j}{|\vec{q}|^2} 
 \biggr) - 
 \frac{1}{2 |\vec{q}|^2}
 \biggl[ 1 - 2 q_0 L(q) + \frac{q_0^2}{q^2} \biggr]
 \biggl( 
  \delta^{ij} - 3 \frac{q^i q^j}{|\vec{q}|^2} 
 \biggr) 
 \;, 
\ea
where $v^\mu \equiv (1,v^i)$,
$q\equiv (q^0,\vec{q})$, our metric convention is ($+$$-$$-$$-$),
and
\be
 L(q) \equiv \frac{1}{2|\vec{q}|} \ln\frac{q_0 + |\vec{q}|}{q_0 - |\vec{q}|}
 \approx 
 - \frac{i \pi}{2|\vec{q}|} + \frac{q_0}{|\vec{q}|^2} + 
 \frac{q_0^3}{3 |\vec{q}|^4} + ...
 \;.  \la{L}
\ee
Integrals with higher powers of $v\cdot q$ in the denominator
can be obtained through the partial derivatives $\partial/\partial q_0$.
The following identities (which can be derived by certain partial 
integrations, or by explicit inspection) are often very useful:
\ba
  \int_v \frac{v^\alpha q^\beta}{(v\cdot q)^2} 
  \epsilon_{\alpha\beta\mu\nu} I^\mu J^\nu
  & = & 
  \int_v \frac{v^\alpha }{v\cdot q}
  \epsilon_{0 \alpha\mu\nu} I^\mu J^\nu
  \;, \la{pi1} \\
  \int_v \frac{v^\alpha v^\beta}{(v\cdot q)^2}
  q_{[\alpha} I_{\mu]} q_{[\beta}J_{\nu]} \eta^{\mu\nu} 
  \!\! & = & \!\!
  2 \int_v \frac{v^\alpha v^\beta}{v\cdot q}
  I_\alpha q_{[\beta}J_{0]} = 
  2 \int_v \frac{v^\alpha v^\beta}{v\cdot q}
  q_{[\alpha} I_{0]} J_\beta 
  \;. \la{pi2}
\ea
Here $q_{[\alpha}I_{\mu]} \equiv q_\alpha I_\mu - q_\mu I_\alpha$, 
and $I,J$ are arbitrary Lorentz vectors.

\newpage



\begin{thebibliography}{99}

\bibitem{wr}
A.N.~Redlich and L.C.R.~Wijewardhana,
Phys.\ Rev.\ Lett.\  {54} (1985) 970;
%
K.~Tsokos,
Phys.\ Lett.\ B {157} (1985) 413.

\bibitem{ns}
A.J.~Niemi and G.W.~Semenoff,
Phys.\ Rev.\ Lett.\  {54} (1985) 2166.

\bibitem{djt}
S.~Deser, R.~Jackiw and S.~Templeton,
Annals Phys.\  {140} (1982) 372. 

\bibitem{gth}
G.~'t Hooft,
Phys.\ Rev.\ Lett.\  {37} (1976) 8.

\bibitem{cdeo}
B.A.~Campbell, S.~Davidson, J.R.~Ellis and K.A.~Olive,
Phys.\ Lett.\ B {297} (1992) 118
[hep-ph/9302221];
%
J.M.~Cline, K.~Kainulainen and K.A.~Olive,
Phys.\ Rev.\ Lett.\  {71} (1993) 2372
[hep-ph/9304321].

\bibitem{js}
M.~Joyce and M.E.~Shaposhnikov,
Phys.\ Rev.\ Lett.\  {79} (1997) 1193
[astro-ph/9703005].

\bibitem{decay}
M.~Christensson, M.~Hindmarsh and A.~Brandenburg,
Astron.\ Nachr.\  {326} (2005) 393
[astro-ph/0209119];
%
L.~Campanelli,
Phys.\ Rev.\ D {70} (2004) 083009
[astro-ph/0407056];
%
R.~Banerjee and K.~Jedamzik,
Phys.\ Rev.\ D {70} (2004) 123003
[astro-ph/0410032].

\bibitem{ewpt}
P.~Elmfors, K.~Enqvist and K.~Kainulainen,
Phys.\ Lett.\ B {440} (1998) 269
[hep-ph/9806403];
%
K.~Kajantie, M.~Laine, J.~Peisa, K.~Rummukainen and M.E.~Shaposhnikov,
Nucl.\ Phys.\ B {544} (1999) 357
[hep-lat/9809004].

\bibitem{gr}
D.~Comelli, D.~Grasso, M.~Pietroni and A.~Riotto,
Phys.\ Lett.\ B {458} (1999) 304
[hep-ph/9903227].

\bibitem{pa}
G.~Piccinelli and A.~Ayala,
Lect.\ Notes Phys.\  {646} (2004) 293
[hep-ph/0404033];
%
L.~Campanelli, P.~Cea, G.L.~Fogli and L.~Tedesco,
astro-ph/0505531.

\bibitem{gs}
M.~Giovannini and M.E.~Shaposhnikov,
Phys.\ Rev.\ Lett.\  {80} (1998) 22
[hep-ph/9708303];
%
Phys.\ Rev.\ D {57} (1998) 2186
[hep-ph/9710234].

\bibitem{kss}
J.B.~Rehm and K.~Jedamzik,
Phys.\ Rev.\ Lett.\  {81} (1998) 3307
[astro-ph/9802255];
%
Phys.\ Rev.\ D {63} (2001) 043509
[astro-ph/0006381];
%
H.~Kurki-Suonio and E.~Sihvola,
Phys.\ Rev.\ Lett.\  {84} (2000) 3756
[astro-ph/9912473];
%
Phys.\ Rev.\ D {62} (2000) 103508
[astro-ph/0006448].

\bibitem{htl}
R.D.~Pisarski,
Phys.\ Rev.\ Lett.\  {63} (1989) 1129;
%
J.~Frenkel and J.C.~Taylor,
Nucl.\ Phys.\ B {334} (1990) 199;
%
E.~Braaten and R.D.~Pisarski,
Nucl.\ Phys.\ B {337} (1990) 569;
%
J.C.~Taylor and S.M.H.~Wong,
Nucl.\ Phys.\ B {346} (1990) 115.

\bibitem{htlaction}
J.~Frenkel and J.C.~Taylor,
Nucl.\ Phys.\ B {374} (1992) 156;
%
E.~Braaten and R.D.~Pisarski,
Phys.\ Rev.\ D {45} (1992) 1827.

\bibitem{dr}
P. Ginsparg, 
Nucl.\ Phys.\ B 170 (1980) 388;
%
T. Appelquist and R.D. Pisarski,
Phys.\ Rev.\ D 23 (1981) 2305.

\bibitem{parity}
K.~Kajantie, M.~Laine, K.~Rummukainen and M.E.~Shaposhnikov,
Phys.\ Lett.\ B {423} (1998) 137
[hep-ph/9710538].

\bibitem{ag}
A.~Gynther,
Phys.\ Rev.\ D {68} (2003) 016001
[hep-ph/0303019].

\bibitem{asy}
P.~Arnold, D.~Son and L.G.~Yaffe,
Phys.\ Rev.\ D {55} (1997) 6264
[hep-ph/9609481].

\bibitem{db}
D.~B\"odeker,
Phys.\ Lett.\ B {426} (1998) 351
[hep-ph/9801430];
%
Nucl.\ Phys.\ B {559} (1999) 502
[hep-ph/9905239].

\bibitem{bmr}
D.~B\"odeker, G.D.~Moore and K.~Rummukainen,
Phys.\ Rev.\ D {61} (2000) 056003
[hep-ph/9907545].

\bibitem{ay}
P.~Arnold and L.G.~Yaffe,
Phys.\ Rev.\ D {62} (2000) 125013
[hep-ph/9912305];
%
D.~B\"odeker,
Nucl.\ Phys.\ B {647} (2002) 512
[hep-ph/0205202].

\bibitem{gdm}
G.D.~Moore,
Phys.\ Rev.\ D {62} (2000) 085011
[hep-ph/0001216];
%
hep-ph/0009161.

\bibitem{krs}
V.A.~Kuzmin, V.A.~Rubakov and M.E.~Shaposhnikov,
Phys.\ Lett.\ B {155} (1985) 36.

\bibitem{htlold}
V.P.~Silin, 
Sov.\ Phys.\ JETP {11} (1960) 1136
[Zh.\ Eksp.\ Teor.\ Fiz.\ {38} (1960) 1577];
%
V.V.~Klimov,
Sov.\ Phys.\ JETP {55} (1982) 199
[Zh.\ Eksp.\ Teor.\ Fiz.\  {82} (1982) 336];
%
H.A.~Weldon,
Phys.\ Rev.\ D {26} (1982) 1394.


\bibitem{htlmu}
D.~B\"odeker and M.~Laine,
JHEP {09} (2001) 029
[hep-ph/0108034].

\bibitem{cm}
M.~Laine and C.~Manuel,
Phys.\ Rev.\ D {65} (2002) 077902
[hep-ph/0111113];
%
C.~Manuel and S.~Mrowczynski,
Phys.\ Rev.\ D {67} (2003) 014015
[hep-ph/0206209].

\bibitem{kin}
J.P.~Blaizot and E.~Iancu,
Phys.\ Rev.\ Lett.\  {70} (1993) 3376
[hep-ph/9301236];
%
P.F.~Kelly, Q.~Liu, C.~Lucchesi and C.~Manuel,
Phys.\ Rev.\ Lett.\  {72} (1994) 3461
[hep-ph/9403403];
%
F.T.~Brandt, J.~Frenkel and J.C.~Taylor,
Nucl.\ Phys.\ B {437} (1995) 433
[hep-th/9411130].

\bibitem{eff}
P.~Arnold, G.D.~Moore and L.G.~Yaffe,
JHEP {01} (2003) 030
[hep-ph/0209353].

\bibitem{var}
V.A.~Rubakov and A.N.~Tavkhelidze,
Phys.\ Lett.\ B {165} (1985) 109;
%
V.A.~Rubakov,
Prog.\ Theor.\ Phys.\  {75} (1986) 366.


\bibitem{amy}
P.~Arnold, G.D.~Moore and L.G.~Yaffe,
JHEP {11} (2000) 001
[hep-ph/0010177];
%
JHEP {05} (2003) 051
[hep-ph/0302165].

\bibitem{ga}
M.A.~Valle Basagoiti,
Phys.\ Rev.\ D {66} (2002) 045005
[hep-ph/0204334];
%
G.~Aarts and J.M.~Mart\'{\i}nez Resco,
JHEP {11} (2002) 022
[hep-ph/0209048].

\bibitem{allorder}
M.~Laine and M.E.~Shaposhnikov,
Phys.\ Lett.\ B {463} (1999) 280
[hep-th/9907194].


\end{thebibliography}
\end{document}